\documentclass[12pt]{article}
\usepackage{amsmath}
\usepackage{latexsym}
\usepackage{epsfig}
\usepackage{graphics}
\title{{\bf Analytical On-shell Calculation of Higher Order Scattering: Massive Particles}}
\author{Barry R. Holstein\\
Department of Physics-LGRT\\
University of Massachusetts\\
Amherst, MA  01003\\
and\\
Kavli Institute for Theoretical Physics\\
University of California\\
Santa Barbara, CA  93016}
\begin{titlepage}
\begin{document}
\maketitle
\begin{abstract}
We demonstrate that the use of on-shell methods, involving calculation of the discontinuity across the t-channel cut associated with the exchange of a pair of massless particles, can be used to evaluate loop contributions to both the electromagnetic and gravitational scattering of massive systems.  In the gravitational case, the use of factorization permits a straightforward and algebraic calculation of higher order scattering results, which were obtained previously by much more arduous Feynman diagram techniques.
\end{abstract}

\end{titlepage}

\section{Introduction}
In classical physics the discussion of Coulomb and of gravitational scattering, involving long range $1/r$ potentials, has long been a staple of the curriculum.  In quantum mechanics this is also the case for electromagnetic scattering, and there exist, in addition to the lowest order amplitudes, higher order but long-range scattering corrections, which have been studied over the years~\cite{Iwa71},\cite{Spr93},\cite{Fei88},\cite{Ros08}.  The leading forms of such modifications are of two types---a "classical" term, independent of $\hbar$ and falling as $1/r^2$, together with a "quantum mechanical" component, falling as $\hbar/mr^3$.  Such corrections to electromagnetic scattering were first studied by Iwasaki~\cite{Iwa71} and by Spruch~\cite{Spr93}, using old-fashioned noncovariant perturbation theory.  A rigorous and detailed evaluation was provided by Feinberg and Sucher using a dispersive methods involving the two-photon t-channel cut diagram~\cite{Fei88}, {\it cf.} Figure 1.  A conventional Feynman diagram calculation was provided by Ross and Holstein~\cite{Ros08}. The Feinberg-Sucher calculation, since it utilizes unitarity to evaluate the discontinuity across the $2\gamma$ cut, is an example of an on-shell procedure and we shall in this paper use a similar but simplified technique, since we are seeking only the long range (power-law fall-off) corrections to the leading Coulomb potential.

Despite the parallels between electromagnetism and gravitation, examination of quantum mechanics texts reveals that (with one exception~\cite{Sca07}) the gravitational interaction is generally not discussed, no doubt due to its use of tensor rather then vector currents and its inherent nonlinearity. However, in the research literature there has been a good deal of study of higher order gravitational scattering~\cite{Don94},\cite{Muz95},\cite{Ham95},\cite{Akh97},\cite{Khr03},\cite{Bje03}.  The technique used in all these works is that of a perturbative Feynman diagram expansion.\footnote{An exception is the paper of Bjerrum-Bohr et al. who use on-shell methods\cite{Bje14}. The difference from our calculation is that they perform the integration to obtain covariant forms, which are then expanded to find the low energy non-analytic pieces.  This requires considerably more effort than the mothods described in our paper.}  The calculation is a challenging one, since not only bubble, triangle, box, and cross box diagrams are involved, as in the electromagnetic analog, but also vertex and bubble diagrams containing the triple-graviton vertex are required. In addition one must deal with vacuum polarization bubbles due both to gravitons and to ghosts.  One indication of the challenge presented by such calculations is that, between the seminal work of Donoghue in 1994~\cite{Don94} and the 2003 calculations of Khriplovich and Kirilin~\cite{Khr03} and Bjerrum-Bohr et al.~\cite{Bje03}, any such work contained numerical errors~\cite{Don94},\cite{Muz95},\cite{Ham95},\cite{Akh97}.  We will show below how, in the case of spinless scattering, both electromagnetic and gravitational, calculations can be performed relatively straightforwardly and painlessly using helicity techniques and the on-shell procedures of Feinberg and Sucher~\cite{Fei88},\cite{Bje14}.  In the gravitational case the evaluation is simplified enormously by the use of factorization, which asserts that the two graviton amplitudes can be written as a product of electromagnetic Compton amplitudes accompanied by a kinematic factor~\cite{Cho95},\cite{Ber02},\cite{Hol06}.

The structure of the paper is as follows:  In section 2 we review the treatment of electromagnetic Compton scattering and show how its use in the two-photon cut diagram yields the t-channel discontinuity and thereby the associated higher order amplitude for the scattering of both charged and neutral systems. In section 3, we perform the parallel calculation for the case of the gravitational interaction, including scattering by a massive polarizable system.  For both electromagnetism and gravity we find complete agreement with previous evaluations, which required considerably more calculational gravitas.  Our results are summarized in a brief concluding section and outlines for future work are described.  Needed angular integrals and connection to recent work are provided in an Appendices.

\section{Electromagnetic Interactions}

We begin with the simple case of electromagnetism and examine the effects that arise when the strictures of quantum mechanics are imposed. Making the connection between the quantum and classical pictures of electrodynamics can be made by noting that the interaction Lagrangian is of the form
\begin{equation}
{\cal L}_{int}=-eA^\mu J_\mu
\end{equation}
where $e=\sqrt{4\pi\alpha_{em}}$ is the electric charge, $A^\mu$ is the electromagnetic vector potential, and $J_\mu$ is the electromagnetic current, which at leading order has the spin-zero matrix element
\begin{equation}
<p_2|J_\mu|p_1>=(p_1+p_2)_\mu+{\cal O}(q)
\end{equation}
where $q=p_1-p_2$. Working in Coulomb gauge---$\boldsymbol{\nabla}\cdot\boldsymbol{A}=0$---and using the Maxwell equation
\begin{equation}
\Box A_\mu=eJ_\mu\, ,
\end{equation}
which has the solution
\begin{equation}
A_\mu(x)=\int {d^4q\over (2\pi)^4}e^{iq\cdot x}{1\over q^2+i\epsilon}J_\mu(q)\, ,
\end{equation}
the interaction between two such systems both having charge $e$ is given by
\begin{equation}
{\rm Amp}_0^{em}(q)=e^2{(p_1+p_2)\cdot(p_3+p_4)\over q^2}+\ldots\, .
\end{equation}
where $q=p_i-p_f$ is the momentum transfer.  In the nonrelativisitic limit
$$q_\mu\stackrel{NR}{\longrightarrow} (0,\boldsymbol{q})$$
so that, including the normalizing factor $1/\sqrt{16E_1E_2E_3E_4}\simeq 1/4m_Am_B$
\begin{equation}
{\cal A}_0^{em}(q)\equiv{1\over \sqrt{16E_1E_2E_3E_4}}{\rm Amp}_0^{em}(q)\stackrel{NR}{\longrightarrow}{e^2\over \boldsymbol{q}^2}
\end{equation}
Making the transition to coordinate space by taking the Fourier transform then yields the familiar Coulomb potential\footnote{The result Eq. (\ref{eq:hg}) follows from taking the inverse Fourier transform of the Born approximation
\begin{equation}
{\rm Amp}(q)=<\boldsymbol{p}_f|\hat{V}|\boldsymbol{p}_i>=\int d^3r e^{i\boldsymbol{q}\cdot\boldsymbol{r}}V(r)
\end{equation}}

\begin{equation}
V_0^{em}(r)=\int{d^3q\over (2\pi)^3}e^{-i\boldsymbol{q}\cdot\boldsymbol{r}}{\cal A}_0^{em}(q)={e^2\over 4\pi r}\label{eq:hg}
\end{equation}

\begin{figure}
\begin{center}
\epsfig{file=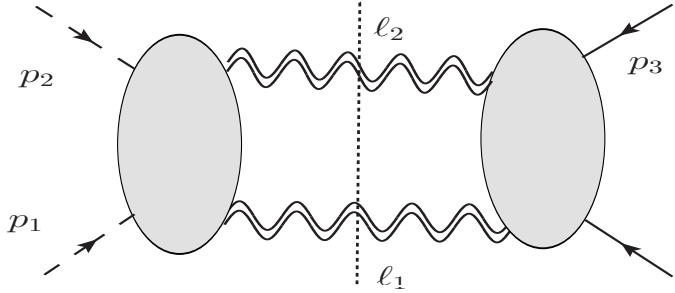,height=4cm,width=10cm}
\caption{\small The two photon cut for the amplitude between a massive
scalar particles. The  grey blobs represent tree-level Compton amplitudes.}
\end{center}
\end{figure}

Of course, there exist higher order modifications of this transition amplitude in quantum mechanics, which take the form of corrections to the interaction potential, and  these are the quantities that we wish to study.  The traditional way to generate these higher order pieces is to use Feynman diagrams.  This procedure is somewhat involved, and details can be found in the literature, as referenced above.  In the present paper we opt for a simpler method to provide what we are seeking, which is the long-range component of these corrections.  The key point in this regard is that, as a function of the invariant momentum transfer $q^2$, the corrections are of two types---analytic and non-analytic.  The former include polynomial functions which multiply the leading $1/q^2$ amplitude.  The result is a polynomial in $q^2$, which, when Fourier-transformed, yields a sum of terms involving $\delta^3(\boldsymbol{r})$ or its derivatives, {\it i.e.,} a {\it local} function with no support away from the origin.  On the other hand, in the case of nonanalytic functions, such as $\sqrt{-q^2}$ or $\log{-q^2}$, the Fourier-transform yields long-distance power law behavior---$1/r^n$ with $n\geq 2$---and it is this type of modification we are seeking.  In order to identify such terms it is sufficient to use the feature that the scattering amplitude is an analytic function of the Mandelstam variables $s,t,u$, with discontinuities prescribed by unitarity.  In order to provide the nonanalytic structure in $t=q^2$ we use the result from S-matrix unitarity that the discontinuity of the scattering amplitude across the right hand cut is given by
\begin{equation}
 {\rm Disc}\, T_{fi}=i(T_{fi}-{T^\dagger}_{fi})=-\sum_n T_{fn}T^\dagger_{ni}\label{eq:mk}
\end{equation}
so that, in the case of elastic scattering, what we require is the product of the Compton creation and annihilation amplitudes---$P+P\rightarrow\gamma+\gamma\rightarrow P'+P'$---{\it cf.} Figure 1.  Considering, for simplicity, spin zero-spin zero scattering, the  t-channel spinless Compton amplitude ({\it i.e.,} the amplitude for two spinless particles of charge $e$ and mass $m_A$ to annihilate into a pair of photons), arises from the diagrams shown in Figure 2 and is easily found~\cite{Hol14}
\begin{equation}
{\rm Amp}_0^A=2e^2\left(\epsilon_1^*\cdot\epsilon_2^*-{\epsilon_1^*\cdot p_1\epsilon_2^*\cdot p_2\over p_1\cdot k_1}-{\epsilon_1^*\cdot p_2\epsilon_2^*\cdot p_1\over p_1\cdot k_2}\right)
\end{equation}

\begin{figure}
\begin{center}
\epsfig{file=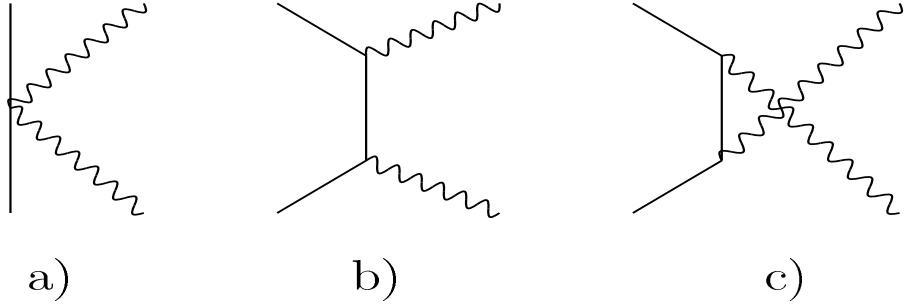,height=4cm,width=12cm}
\end{center}
\caption{{Shown are the a) seagull, b) Born, and c) Cross-Born diagrams contributing to Compton scattering. Here the solid lines represent massive
scalars while the wiggly lines are photons.}}\label{fig:compton}
\end{figure}

It is convenient to use the helicity formalism~\cite{Jac59}, where helicity is defined as the projection of the photon spin on the momentum axis.  The helicity amplitudes for t-channel spin-0-spin-0 Compton annihilation are found in the center of mass frame to have the form
\begin{eqnarray}
^AA_0^{EM}(++)&=&^AA_0^{EM}(--)=2e^2\left({{m_A^2\over E_A^2-\boldsymbol{p}_A^2\cos^2\theta_A}}\right)\,,\nonumber\\
^AA_0^{EM}(+-)&=&^AA_0^{EM}(-+)=2e^2\left({\boldsymbol{p}_A^2\sin^2\theta_A\over E_A^2-\boldsymbol{p}_A^2\cos^2\theta_A}\right)\,,
\end{eqnarray}
where $m_A,\,E_A,\,\pm\boldsymbol{p}_A$ are the mass, energy, momentum of the spinless particles and $\theta_A$ is the scattering angle, {\it i.e.} the angle of the outgoing photons---$\cos\theta_A=\hat{\boldsymbol{p}}_A\cdot\hat{\boldsymbol{k}}$.   It was shown by Feinberg and Sucher that the annihilation amplitudes $A+A'\rightarrow\gamma_1+\gamma_2$ needed in the unitarity relation---Eq. (\ref{eq:mk})---can be generated by making an analytic continuation to imaginary momentum $\boldsymbol{p}_A\rightarrow im_A\xi_A\hat{\boldsymbol{p}}_A$, where $\xi_A^2=1-{t\over 4m_A^2}$ and $t=(p_A-p_A')^2$ is the momentum transfer~\cite{Fei88}.  Then
\begin{eqnarray}
^AA_0^{EM}(++)&=&^AA_0^{EM}(--)= 2e^2{1+\tau_A^2\over d_A}\,,\nonumber\\
^AA_0^{EM}(+-)&=&^AA_0^{EM}(-+)= 2e^2{1-x_A^2\over d_A}\,,\label{eq:bv}
\end{eqnarray}
where we have defined $\tau_A=\sqrt{t}/2m_A\xi_A$, $x_A=\hat{\boldsymbol{p}}_A\cdot\hat{\boldsymbol{k}}$, and $d_A=\tau_A^2+x_A^2$.
Equivalently Eq. (\ref{eq:bv}) can be represented succinctly via
\begin{equation}
^AA_0^{EM}(ab)=2e^2{\cal O_A}^{ij}{\epsilon_{1i}^{a*}}{\epsilon_{2j}^{b*}}\end{equation}
where
\begin{equation}
{\cal O}_A^{ij}={1\over d_A}\left(d_A\delta^{ij}+2\hat{p}_A^i\hat{p}_A^j\right)
\end{equation}
Similarly in the outgoing channel $\gamma_1+\gamma_2\rightarrow B+B'$, again for a spinless particle of charge $e$ but now with mass $m_B$, we define the corresponding quantities $\tau_B=\sqrt{t}/2m_B\xi_B$, $x_B=\hat{\boldsymbol{p}}_B\cdot\hat{\boldsymbol{k}}$, and $d_B=\tau_B^2+x_B^2$ and the helicity amplitudes can be described by
\begin{equation}
^BA^{em}_0(cd)=2e^2\epsilon_{1k}^{c}\epsilon_{2\ell}^{d}{\cal O_B}^{*k\ell}\end{equation}
where
\begin{equation}
{\cal O}_B^{k\ell}={1\over d_B}\left(d_B\delta^{k\ell}+2\hat{p}_B^k\hat{p}_B^\ell\right)
\end{equation}
Substituting in Eq. (\ref{eq:mk}), we determine the discontinuity for scattering of spinless particles having masses $m_A,m_B$ across the t-channel two-photon cut for scattering in the CM frame, {\it cf.} Figure 1\footnote{Note that we have divided by the normalizing factor $4E_AE_B\simeq 4m_Am_B$ since we will be using this amplitude in the nonrelativistic limit.}
\begin{eqnarray}
{\rm Disc}\,{\rm Amp}_2^{em}(q)&=&-{i\over 2!}{(2e^2)^2\over 4m_Am_B}\int {d^3k_1\over (2\pi)^32k_{10}}{d^3k_2\over (2\pi)^32k_{20}}(2\pi)^4\delta^4(p_1+p_2-k_1-k_2)\nonumber\\
&\times&\sum_{a,b=1}^2\left[{\cal O}_A^{ij}\epsilon^{a*}_{1i}\epsilon^{b*}_{2j}\epsilon^a_{1k}\epsilon^b_{2\ell}{\cal O}_B^{k\ell*}\right]\nonumber\\
&=&-i{e^4\over 16\pi m_Am_B}<\sum_{i,j,k,\ell=1}^3{\cal O}_A^{ij}\delta^T_{ik}\delta^T_{j\ell}{\cal O}_B^{k\ell*}>\nonumber\\
\quad
\end{eqnarray}
where
\begin{equation}
\delta^T_{ik}=\sum_{a=1}^2\epsilon_i^{a*}\epsilon_k^{a}=\delta_{ik}-\hat{k}_i\hat{k}_k
\end{equation}
represents the sum over photon polarizations and
$$<G>\equiv \int{d\Omega_{\hat\,{\boldsymbol{k}}}\over 4\pi}G$$
defines the average over solid-angle.  Performing the polarization sums, we find
\begin{eqnarray}
<\sum_{i,j,k,\ell=1}^3{\cal O}_A^{ij}\delta^T_{ik}\delta^T_{j\ell}{\cal O}_B^{k\ell*}>&=&<{1\over d_Ad_B}\left(4(y-x_Ax_B)^2-2(1-x_A^2)(1-x_B^2)\right.\nonumber\\
&+&\left.2(1+\tau_A^2)(1+\tau_B^2)\right)>\label{eq:pj}
\end{eqnarray}
where
\begin{equation}
y=\hat{\boldsymbol{p}}_A\cdot\hat{\boldsymbol{p}}_B={2s+t-2m_A^2-2m_B^2\over 4m_A\xi_Am_B\xi_B}
\end{equation}
characterizes the angle between incoming and outgoing spinless particles.  Eq. (\ref{eq:pj}) is exact, but since we are seeking the long-range behavior of the scattering amplitude, we need only the small-$t$ dependence whereby we have
\begin{equation}
 <\sum_{i,j,k,\ell=1}^3{\cal O}_A^{ij}\delta^T_{ik}\delta^T_{j\ell}{\cal O}_B^{k\ell*}> \stackrel{t<<m_A^2,m_B^2}{\longrightarrow} <{1\over d_Ad_B}\left(4(y-x_Ax_B)^2+2x_A^2+2x_B^2-2x_A^2x_B^2\right)>
\end{equation}
For small $t$ and at threshold---$s\rightarrow s_0=(m_A+m_B)^2$---
\begin{equation}
y(s_0,t)={2s_0+t-2m_A^2-2m_B^2\over 4m_A\xi_Am_B\xi_B}\stackrel{t\rightarrow 0}{\longrightarrow} 1+{\cal O}(t)\,.
\end{equation}
so
\begin{eqnarray}
&&{\rm Disc}\,{\rm Amp}_2^{em}(q)\simeq -i{e^4\over 8\pi m_Am_B}<{1\over d_Ad_B}\left(2y^2-4yx_Ax_B+x_A^2+x_B^2+x_A^2x_B^2\right)>\nonumber\\
&&\stackrel{y\rightarrow 1}{\longrightarrow}-i{e^4\over 8\pi m_Am_B}<{1\over d_Ad_B}\left(2-4x_Ax_B+x_A^2+x_B^2+x_A^2x_B^2\right)>.\nonumber\\
\quad\label{eq:gb}
\end{eqnarray}
The needed angular integrals
$$I_{mn}=<{x_A^mx_B^n\over d_Ad_B}>$$
have been given by Feinberg and Sucher and are quoted in Appendix B~\cite{Fei70}, yielding
\begin{eqnarray}
{\rm Disc}\,{\rm Amp}_2^{em}(q)&\simeq&-i{e^4\over 8\pi m_Am_B}\left[2I_{00}-4I_{11}+I_{20}+I_{02}+I_{22}\right]\nonumber\\
&=&-i{e^4\over 8\pi m_Am_B}\left[2\left(-{1\over 3}+i4\pi{m_rm_Am_B\over p_0t}\right)-4\left(-1+i4\pi{m_r\over p_0}\right)\right.\nonumber\\
&+&\left.\left({\pi m_A\over \sqrt{t}}-1\right)
+\left({\pi m_B\over \sqrt{t}}-1\right)+1\right]\nonumber\\
&=&-i{e^4\over 8\pi m_Am_B}\left[{\pi(m_A+m_B)\over \sqrt{t}}+{7\over 3}+i{4\pi m_Am_Bm_r\over p_0t}+\ldots\right]\nonumber\\
\quad
\end{eqnarray}
where $p_0=\sqrt{{m_r(s-s_0)\over m_A+m_B}}$ is the center of mass momentum for the spinless scattering process and $m_r=m_Am_B/(m_A+m_B)$ is the reduced mass. Since
\begin{equation}
{\rm Disc}\,\left[\log(-t),\,\sqrt{1\over -t}\right]=\left[2\pi i,\,-i{2\pi^2\over \sqrt{t}}\right]
\end{equation}
the scattering amplitude is
\begin{equation}
{\rm Amp}_2^{em}(q)=-{e^4\over 16\pi^2 m_Am_B}\left[{7\over 3}L-S(m_A+m_B)+4\pi i{m_rm_Am_B\over p_0t}L+\ldots\right]\,,\label{eq:vc}
\end{equation}
where we have defined $L=\log(-t)$ and $S=\pi^2/\sqrt{-t}$.  The imaginary component of Eq. (\ref{eq:vc}) represents the Coulomb phase or equivalently the contribution of the second Born approximation, which must be subtracted in order to define a proper higher-order potential.  Using~\cite{Dal51}
\begin{eqnarray}
B_2^{em}(q)&=&i\int{d^3\ell\over (2\pi)^3}{e^2\over |\boldsymbol{p}_f-\boldsymbol{\ell}|^2+\lambda^2}
{i\over {p_0^2\over 2m_r}-{\ell^2\over 2m_r}+i\epsilon}{e^2\over |\boldsymbol{\ell}-\boldsymbol{p}_i|^2+\lambda^2}\nonumber\\
&=&-i{e^4\over 4\pi^2}{m_r\over p_0}{\log(-t)\over t}\label{eq:cx}
\end{eqnarray}
what remains is the higher order electromagnetic amplitude we are seeking.  Writing $t=q^2$ and taking the nonrelativistic limit, we find the second-order effective potential
\begin{eqnarray}
V_2^{em}(r)&=&\int{d^3q\over (2\pi)^3}e^{-i\boldsymbol{q}\cdot\boldsymbol{r}}\left({\rm Amp}_2^{em}(q)-B_2^{em}(q)\right)\,,\nonumber\\
&=&-{\alpha_{em}^2(m_A+m_B)\over 2m_Am_Br^2}-{7\alpha_{em}^2\hbar\over 6\pi m_Am_B r^3}\label{eq:dz}
\end{eqnarray}
In comparing Eq. (\ref{eq:dz}) with previous calculations, it is necessary to understand an important point made by Sucher~\cite{Suc94}, which is that the result for the {\it classical} potential depends on the specific form of the lowest order potential and the propagator used to generate the Born subtraction. The result Eq. (\ref{eq:cx}) follows from use of the simplest nonrelativistic forms for each. Inclusion of relativistic corrections in either the potential or the propagator (or both) will yield the same imaginary piece as found above but also, in general, a term involving $\pi^2\over \sqrt{-t}$, which generates a correction to the classical potential, with the quantum piece remaining unchanged.  Consequently, the result quoted in Eq. (\ref{eq:dz}) agrees with the previous form found by Ross and Holstein~\cite{Ros08} but {\it not} with the results of Feinberg and Sucher, of Iwasaki, or of Spruch.  What {\it is} the same in each calculation is the form of the on-shell scattering amplitude
\begin{equation}
{\rm Amp}_2^{em}(q)=\int d^3r e^{-i\boldsymbol{q}\cdot\boldsymbol{r}}(V_0^{em}(r)+V_2^{em}(r))+B_2^{em}(q)
\end{equation}
Although identical results are obtained from either procedure, as shown in Appendix A, the Feinberg-Sucher technique is calculationally simpler than the conventional Feynman diagram method, since the latter involves evaluation of the separate contributions from the bubble, triangle, box and cross-box diagrams shown in Figure 3.  This simplification will be found to be even more significant in the case of gravitational scattering, as discussed in the next section.

\begin{figure}
\begin{center}
\epsfig{file=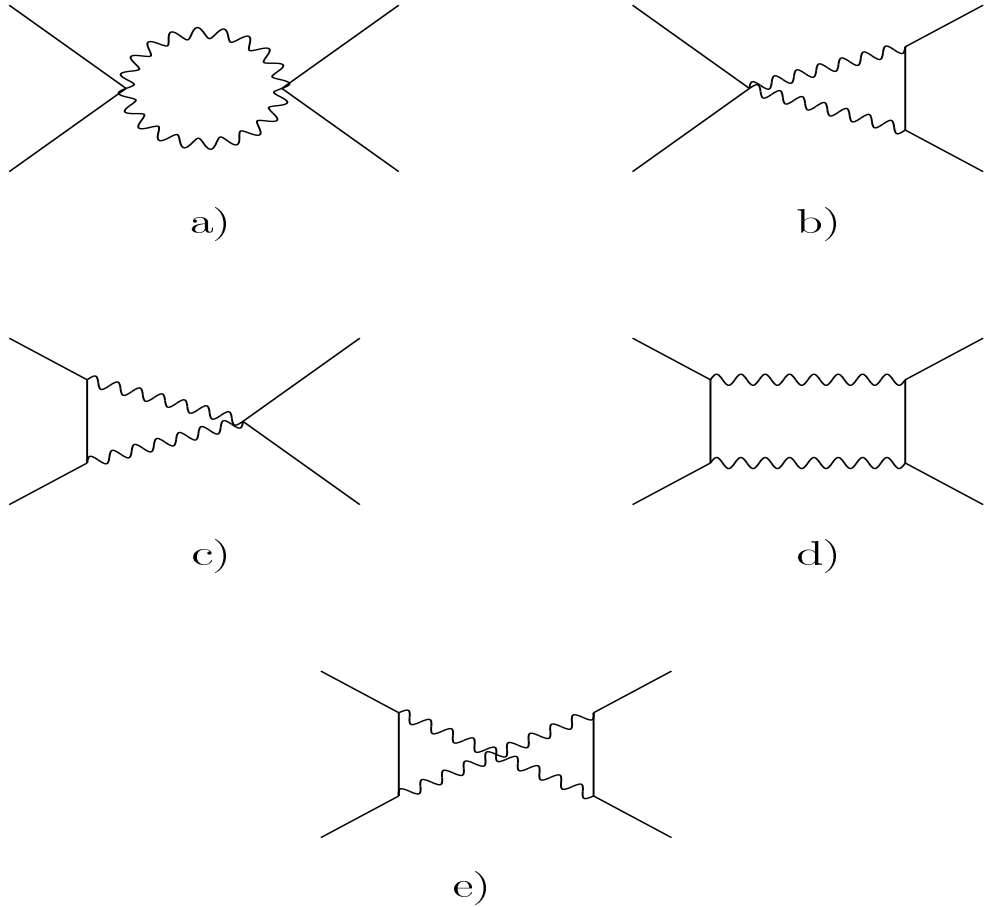,height=12cm,width=13cm}
\caption{{Shown are the a) bubble, b),c) triangle, d) Box and e) Cross-box diagrams contributing to spinless particle scattering.  Here the solid lines designate the massive spinless particles, while the wiggly lines represent photons in the case of the electromagnetic interaction or gravitons in the case of gravitational scattering.}}
\end{center}
\end{figure}

\subsection{Polarizable Electromagnetic Scattering}

However, before proceeding to our analysis of gravitational scattering, we note that it is also
straightforward and interesting to evaluate the electromagnetic interaction of a charged particle
with a {\it neutral} spinless system, wherein the source of a photon pair is characterized by
electric and magnetic polarizabilities $\alpha_E$ and $\beta_M$, defined by the effective Hamiltonian
\begin{equation}
H_{eff}=-{1\over 2}(4\pi\alpha_E\boldsymbol{E}^2+4\pi\beta_M\boldsymbol{H}^2)
\end{equation}
For a spinless system with mass $m_A$, writing the electric and magnetic fields in terms of the field tensor $F_{\mu\nu}$ via
\begin{eqnarray}
\boldsymbol{E}_i&=&{1\over m_A}p_1^\alpha F_{\alpha i}\nonumber\\
\boldsymbol{B}_i&=&\epsilon_{i\alpha\beta\gamma}F^{\alpha\beta}p_1^\gamma{1\over 2m_A}
\end{eqnarray}
the Hamiltonian becomes
\begin{equation}
H=-{4\pi\alpha_E\over 2m_A^2}p_1^\alpha F_{\alpha\beta}F^{\beta\gamma}p_{1\gamma}-{4\pi\beta_M\over 8m_A^2}\epsilon^{\alpha\beta\gamma\delta}
{\epsilon_\alpha}^{\rho\sigma\lambda}F_{\rho\sigma}p_{1\lambda}
\end{equation}
The electric and magnetic transition amplitudes for annihilation into a photon pair are then
\begin{eqnarray}
{\rm Amp}_E&=&-4\pi\alpha_E\epsilon_{1\mu}^*\epsilon_{2\nu}^*T_E^{\mu\nu}(p_1,k_1,k_2)\nonumber\\
{\rm Amp}_M&=&-4\pi\beta_M\epsilon_{1\mu}^*\epsilon_{2\nu}^*T_M^{\mu\nu}(p_1,k_1,k_2)
\end{eqnarray}
with
\begin{eqnarray}
T_E^{\mu\nu}(p_1,k_1,k_2)&=&{1\over m_A^2}\left(\eta^{\mu\nu}p_1\cdot k_1p_1\cdot k_2+p_1^\mu p_1^\nu k_1\cdot k_2-p_1^\mu k_1^\nu p_1\cdot k_2-k_2^\mu p_1^\nu p_1\cdot k_1\right)\nonumber\\
T_M^{\mu\nu}(p_1,k_1,k_2)&=&T_E^{\mu\nu}(p_1,k_1,k_2)-\eta^{\mu\nu}k_1\cdot k_2+k_2^\mu k_1^\nu
\end{eqnarray}
The corresponding t-channel helicity amplitudes are, after Feinberg-Sucher continuation, found to be
\begin{eqnarray}
N_A(++)&=&N_A(--)\stackrel{FS}{\longrightarrow}\pi t(\alpha_E^A-\beta_M^A)\nonumber\\
N_A(+-)&=&N_A(-+)\stackrel{FS}{\longrightarrow}\pi t(\alpha_E^A+\beta_M^A)(1-x_A^2)
\end{eqnarray}
{\it i.e.,}
\begin{equation}
N_A^{ab}=\pi t{\cal V}_A^{ij}\epsilon^{a*}_{1i}\epsilon^{b*}_{2j}
\end{equation}
with
\begin{equation}
{\cal V}_A^{ij}=(\alpha_E^Ax_A^2-\beta_M^A(2-x_A^2))\delta^{ij}+2(\alpha_E^A+\beta_M^A)\hat{p}_A^i\hat{p}_A^j
\end{equation}
We find then the discontinuity for spinless charged-neutral scattering to be
\begin{eqnarray}
&&{\rm Disc}\,{\rm Amp}_N=-{i\over 2!}\int {d^3k_1\over (2\pi)^32k_1^0}{d^3k_2\over (2\pi)^32k_2^0}(2\pi)^4\delta^4(p_1+p_2-k_1-k_2)\nonumber\\
&\times&\sum_{a,b=1}^2{\pi e^2t\over m_B}{\cal V}_A^{ij}\epsilon^{a*}_{1i}\epsilon^{b*}_{2j}\epsilon^a_{1k}\epsilon^b_{2\ell}{\cal O}_B^{k\ell*}={-ie^2t\over 16m_B}
\sum_{i,j,k,\ell=1}^3<\left[{\cal V}_A^{ij}P^T_{ik}P^T_{j\ell}{\cal O}_B^{k\ell*}\right]>\nonumber\\
&\stackrel{FS}{\longrightarrow}& -{i\pi\alpha_{em} t\over 2m_B}\left[-2\beta_M^A<{1\over d_B}>+(\alpha_E^A+\beta_M^A)<{x_A^2+x_B^2-x_A^2x_B^2+2(y-x_Ax_B)^2\over d_B}>\right]\nonumber\\
\quad
\end{eqnarray}
The needed angular integrals
$$J^B_{mn}=<{x_A^mx_B^n\over d_B}>$$
have been given by Feinberg and Sucher and are quoted in Appendix B~\cite{Fei70}, and we determine
\begin{eqnarray}
{\rm Disc}\,{\rm Amp}_N &\stackrel{y\rightarrow 1}{\longrightarrow}& -2\pi i{\alpha_{em} t\over 4m_B}\left[2\alpha_E^AJ^B_{00}+(\alpha_E^A+\beta_M^A)\left(J^B_{02}+J^B_{20}-4J^B_{11}+J^B_{22}\right)\right]\nonumber\\
&=& -2\pi i{\alpha_{em} t\over 4m_B}\left[({2\pi m_B\over \sqrt{-t}}-{11\over 3})\alpha_E^A-{5\over 3}\beta_M^A\right]
\end{eqnarray}
which corresponds to the scattering amplitude
\begin{equation}
{\rm Amp}_N(q)=\alpha_{em}\alpha_E^A{\pi^2\sqrt{-t}\over 2}+{1\over 3}\alpha_{em}(11\alpha_E^A+5\beta_M^A)t\log -t
\end{equation}
Taking the nonrelativistic limit and Fourier transform, we find the effective potential describing interaction of a charged and neutral system
\begin{eqnarray}
V_N(r)&=&\int {d^3q\over (2\pi)^3}e^{-i\boldsymbol{q}\cdot\boldsymbol{r}}{\rm Amp}_N(q)\nonumber\\
&=&-{\alpha_{em}\alpha_E^A\over 2r^4}+\alpha_{em}{(11\alpha_E^A+5\beta_M^A)\hbar\over 4\pi m_B r^5}
\end{eqnarray}
The form of the classical $\sim 1/r^4$ component of the potential is easily derived.  Suppose a particle of charge $e$ sits at the origin.  Then at location $\boldsymbol{r}$ an electric field $\boldsymbol{E}(\boldsymbol{r})=e\hat{\boldsymbol{r}}/4\pi r^2$ exists.  If there is a neutral particle at this location, there will be an induced electric dipole moment $\boldsymbol{d}_E=4\pi\alpha_E\boldsymbol{E}$ and the corresponding interaction energy is
\begin{equation}
\Delta U=-{1\over 2}\boldsymbol{d}_E\cdot\boldsymbol{E}(\boldsymbol{r})=-{1\over 2}4\pi\alpha_E\boldsymbol{E}^2(\boldsymbol{r})=-{\alpha_{em}\alpha_E\over 2r^4}
\end{equation}
The quantum component was first found by Bernabeu and Tarrach using dispersive methods~\cite{Ber76} and was later reexamined by Feinberg and Sucher, using a generalization of their charged particle interaction calculation~\cite{Fei92}.  Qualitatively it can be considered to arise from zitterbewegung.  That is, classically the potential is well defined as a function of the separation $r$.  However, in quantum mechanics, the location of a particle is uncertain by an amount of order its Compton wavelength $\delta r\sim \hbar/m$.  This argument leads to the replacement
$$V(r)\sim{1\over r^4}\longrightarrow {1\over (r\pm \delta r)^4}\sim {1\over r^4}\pm 4{\hbar\over mr^5} $$
which is the form found above.

We can also examine the interaction of two spinless systems, both of which are characterized by polarizabilities~\cite{Hol14a}.  Then
\begin{eqnarray}
&&{\rm Disc}\,{\rm Amp}_{NN}(q)=-{i\over 2!}\int {d^3k_1\over (2\pi)^32k_1^0}{d^3k_2\over (2\pi)^32k_2^0}(2\pi)^4\delta^4(p_1+p_2-k_1-k_2)\nonumber\\
&\times&\sum_{a,b=1}^2\pi^2t^2{\cal V}_A^{ij}\epsilon^{a*}_{1i}\epsilon^{b*}_{2j}\epsilon^a_{1k}\epsilon^b_{2\ell}{\cal V}_B^{k\ell*}=-i{\pi t^2\over 16}\sum_{i,j,k,\ell=1}^3<\left[{\cal V}_A^{ij}P^T_{ik}P^T_{j\ell}{\cal V}_B^{k\ell*}\right]>\nonumber\\
&\stackrel{FS}{\longrightarrow}&-i2\pi{t^2\over 16}<(\alpha_E^A\alpha_E^B+\beta_M^A\beta_M^B)(2+x_A^2+x_B^2-4x_Ax_B+x_A^2x_B^2)\nonumber\\
&+&(\alpha_E^A\beta_M^B+\alpha_E^B\beta_M^A)(x_A^2+x_B^2-4x_Ax_B+x_A^2x_B^2)>
\end{eqnarray}
The needed angular integrals needed in this case
$$<K_{mn}>=<x_A^mx_B^n>$$ are given in Appendix B, and we find
\begin{eqnarray}
{\rm Disc}\,{\rm Amp}_{NN}(q)&=&-i2\pi{t^2\over 16}\left[(\alpha_E^A\alpha_E^B+\beta_M^A\beta_M^B)(2K_{00}+K_{20}+K_{02}-4K_{11}+K_{22})\right.\nonumber\\
&+&\left.(\alpha_E^A\beta_M^B+\alpha_E^B\beta_M^A)(K_{20}+K_{02}-4K_{11}+K_{22})\right]\nonumber\\
&=&-i2\pi{t^2\over 16}\left[(\alpha_E^A\alpha_E^B+\beta_M^A\beta_M^B){23\over 15}-(\alpha_E^A\beta_M^B+\alpha_E^B\beta_M^A){7\over 15}\right]\nonumber\\
\quad
\end{eqnarray}
which corresponds to the scattering amplitude
\begin{equation}
{\rm Amp}_{NN}(q)={t^2L\over 16}\left[(\alpha_E^A\alpha_E^B+\beta_M^A\beta_M^B){23\over 15}-(\alpha_E^A\beta_M^B+\alpha_E^B\beta_M^A){7\over 15}\right]
\end{equation}
Taking the Fourier transform we find the effective potential
\begin{eqnarray}
V_{NN}(r)&=&\int {d^3q\over (2\pi)^3}e^{-i\boldsymbol{q}\cdot\boldsymbol{r}}{\rm Amp}_{NN}(q)\nonumber\\
&=&{-23(\alpha_E^A\alpha_E^B+\beta_M^A\beta_M^B)+7(\alpha_E^A\beta_M^B+\alpha_E^B\beta_M^A)\over 4\pi r^7}\label{eq:rx}
\end{eqnarray}
which is the familiar Casimir-Polder potential~\cite{Cas48},\cite{Hol01}.  Eq. (\ref{eq:rx}) represents the very long distance form of the van der Waals interaction which at closer distances, where retardation effects are not relevant, has the familiar $1/r^6$ fall-off.

It is interesting to note here that the electric transition helicity amplitudes are proportional to the corresponding point particle charged amplitudes
\begin{equation}
{\rm Amp}_E(ij)=-4\pi\alpha_E\times{d_A\over 2e^2}A_0^A(ij)
\end{equation}
which implies that the corresponding amplitudes for charged particle-charged particle, charged particle-neutral particle, and neutral particle-neutral particle calculations have a simple relation between their angular averages
\begin{eqnarray}
{\rm Disc}\,{\rm Amp}(\rm charged-charged)&=&-i{e^4\over 2\pi}<{1\over d_Ad_B}(2-4x_Ax_B+x_A^2+x_B^2+x_A^2x_B^2)>\nonumber\\
&=&-i{e^4\over 2\pi}(2I_{00}-4I_{11}+I_{20}+I_{02}+I_{22})\nonumber\\
{\rm Disc}\,{\rm Amp}(\rm neutral-charged)&=&-i{e^4\over 2\pi}\times -{4\pi\alpha_E^A\over 2e^2}\nonumber\\
&\times&<{1\over d_B}(2-4x_Ax_B+x_A^2+x_B^2+x_A^2x_B^2)>\nonumber\\
&=&ie^2\alpha_E^A(2J^B_{00}-J^B_{11}+J^B_{20}+J^B_{02}+J^B_{22})\nonumber\\
{\rm Disc}\,{\rm Amp}(\rm neutral-neutral)&=&ie^2\alpha_E^A\times \left(-{4\pi\alpha_E^B\over 2e^2}\right)\nonumber\\ &\times&<(2-4x_Ax_B+x_A^2+x_B^2+x_A^2x_B^2)>\nonumber\\
&=&-i2\pi \alpha_E^A\alpha_E^B(2K_{00}-4K_{11}+K_{20}+K_{02}+K_{22})\,,\nonumber\\
\quad
\end{eqnarray}
{\it i.e.}, they can be obtained from one another by the simple substitution $I_{nm}\rightarrow J^B_{nm}\rightarrow K_{nm}$.  This parallel will be found also in the gravitational case.

\section{Gravitational Scattering}

The case of gravitational scattering of spinless systems can be treated in parallel with the electromagnetic case.  In this case the basic interaction Lagrangian is
\begin{equation}
{\cal L}_{int}={\kappa\over 2}h_{\mu\nu} T^{\mu\nu}
\end{equation}
where $\kappa=\sqrt{32\pi G}$ is the gravitational coupling, $g_{\mu\nu}\equiv\eta_{\mu\nu}+h_{\mu\nu}$ is the metric tensor, and $T^{\mu\nu}$ is the energy-momentum tensor, which at leading order has the spin-zero matrix element
\begin{equation}
<p_2|T^{\mu\nu}(x)|p_1>=2(p_1+p_2)_\mu(p_1+p_2)_\nu e^{i(p_2-p_1)\cdot x}+{\cal O}(q)
\end{equation}
Working in harmonic (or DeDonder) gauge---$\partial^\mu h_{\mu\nu}={1\over 2}\partial_\nu h^\mu_\mu$---and using the (linearized) Einstein equation
\begin{equation}
\Box h_{\mu\nu}=-{1\over 2}\kappa\left(T_{\mu\nu}-{1\over 2}\eta_{\mu\nu}{T^\alpha}_\alpha\right)
\end{equation}
we have the solution for a point mass $m_i$
\begin{equation}
h_{\mu\nu}(x)=\int {d^4q\over (2\pi)^4}e^{iq\cdot x}{\kappa m_i\delta_{\mu\nu}\over 4q^2+i\epsilon}
\end{equation}
so that the interaction between systems with masses $m_A,m_B$ is given by
\begin{equation}
{\rm Amp}_0^{grav}(q)=-{\kappa^2\over 4}{(s-m_A^2-m_B^2+{1\over 2}q^2)^2-2m_A^2m_B^2-{1\over 4} q^4\over q^2}+\ldots
\end{equation}
where $q=p_1-p_2=p_4-p_3$, so that in the nonrelativisitic limit
\begin{equation}
{\cal A}_0^{grav}(q)\equiv{1\over \sqrt{16E_1E_2E_3E_4}}{\rm Amp}_0(q)\stackrel{NR}{\longrightarrow}-4\pi G{m_Am_B\over \boldsymbol{q}^2}
\end{equation}
We can make the transition to coordinate space by taking the Fourier transform, which yields the familiar Newtonian potential
\begin{equation}
V_0^{grav}(r)=\int{d^3q\over (2\pi)^3}e^{-i\boldsymbol{q}\cdot\boldsymbol{r}}{\cal A}^{grav}_0(q)=-G{m_1m_2\over r}\,.
\end{equation}

\begin{figure}
\begin{center}
\epsfig{file=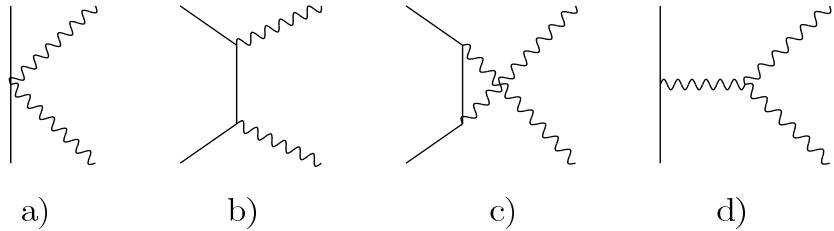,height=3cm,width=11cm}
\caption{{Shown are the a) seagull, b) Born, c) Cross-Born, and d) graviton pole diagrams contributing the gravitational Compton scattering.  Here the solid lines represent massive spinless particles while the wiggly lines are gravitons.}}
\label{fig:gravcomp}
\end{center}
\end{figure}

As in the electromagnetic case, there exist higher order corrections to the transition amplitude generated by quantum mechanics which provide corrections to the lowest order Newtonian form.  We use the unitarity condition Eq. (\ref{eq:mk}), but we now require the discontinuity across the two-{\it graviton} t-channel cut and the result will involve the product of {\it gravitational} Compton scattering amplitudes.  For simplicity, we consider spin zero-spin zero scattering, for which the sum of the four diagrams shown in Figure 4 for spinless gravitational Compton amplitude yield the rather complex form
\begin{eqnarray}
{\rm Amp}_{GC}(S=0)&=&\kappa^2 \left\{-{(\epsilon_i\cdot
p_i)^2(\epsilon_f^*\cdot p_f)^2\over p_i\cdot k_i}+{(\epsilon_f^*\cdot
p_i)^2(\epsilon_i\cdot
p_f)^2\over p_i\cdot k_f}\right.\nonumber\\
&+&\left.\left[\epsilon_f^*\cdot\epsilon_i(\epsilon_i\cdot
p_i\epsilon_f^*\cdot p_f+\epsilon_i\cdot p_f\epsilon_f^*\cdot
p_i)-{1\over 2}k_i\cdot
k_f(\epsilon_f^*\cdot\epsilon_i)^2\right]\right.\nonumber\\
&+&\left.{1\over 2k_i\cdot k_f}
\left[\epsilon_f^*\cdot p_f\epsilon_f^*\cdot
p_i(\epsilon_i\cdot(p_i-p_f))^2\right.\right.\nonumber\\
&+&\left.\left.\epsilon_i\cdot p_i\epsilon_i\cdot
p_f(\epsilon_f^*\cdot(p_i-p_f))^2\right.\right.\nonumber\\
&+&\left.\left.\epsilon_i\cdot(p_i-p_f)\epsilon_f^*\cdot(p_f-p_i)(\epsilon_f^*\cdot
p_f\epsilon_i\cdot p_i+\epsilon_f^*\cdot p_i\epsilon_i\cdot
p_f)\right.\right.\nonumber\\
&-&\left.\left.\epsilon_f^*\cdot\epsilon_i\left(\epsilon_i\cdot(p_i-p_f)\epsilon_f^*\cdot
(p_f-p_i)(p_i\cdot p_f-m^2)\right.\right.\right.\nonumber\\
&+&\left.\left.\left.k_i\cdot k_f(\epsilon_f^*\cdot p_f\epsilon_i\cdot
p_i+\epsilon_f^*\cdot p_i\epsilon_i\cdot p_f)\right.\right.\right.\nonumber\\
&+&\left.\left.\left.\epsilon_i\cdot(p_i-p_f)(\epsilon_f^*\cdot p_fp_i\cdot
k_f+\epsilon_f^*\cdot p_ip_f\cdot k_f)\right.\right.\right.\nonumber\\
&+&\left.\left.\left.\epsilon_f^*\cdot(p_f-p_i)(\epsilon_i\cdot
p_ip_f\cdot
k_i+\epsilon_i\cdot p_fp_i\cdot k_i)\right)\right.\right.\nonumber\\
&+&\left.\left.(\epsilon_f^*\cdot\epsilon_i)^2\left(p_i\cdot k_ip_f\cdot
k_i+p_i\cdot k_fp_f\cdot k_f\right.\right.\right.\nonumber\\
&-&\left.\left.\left.{1\over 2}(p_i\cdot k_ip_f\cdot
k_f+p_i\cdot k_fp_f\cdot k_i)\right.\right.\right.\nonumber\\
&+&\left.\left.\left.{3\over 2}k_i\cdot k_f(p_i\cdot
p_f-m^2)\right)\right]\right\}\label{eq:gx}
\end{eqnarray}
However, Eq. (\ref{eq:gx}) can be greatly simplified, since it has been pointed out that such graviton Compton amplitudes
factorize into a products of two electromagnetic Compton amplitudes multiplied by a simple kinematic
factor~\cite{Cho95},\cite{Ber02}
\begin{equation}
F={p_i\cdot k_ip_i\cdot k_f\over k_i\cdot k_f}\,,
\end{equation}
which, in the center of mass frame has the form
\begin{equation}
F={E^2(E-|\boldsymbol{p}|\cos\theta)(E+|\boldsymbol{p}|\cos\theta)\over 2E^2}={1\over 2}(E^2-\boldsymbol{p}^2\cos^2\theta)
\end{equation}
That is, we have the remarkable identity
\begin{eqnarray}
{\rm Amp}_{C}^{grav}(S=0)&=&{\kappa^2\over 8e^4}F[{\rm Amp}^{em}_C(S=0)]^2\nonumber\\
&=&{\kappa^2\over 2}\left({p_i\cdot k_ip_f\cdot k_f\over k_i\cdot k_f}\right)\left({\epsilon_i\cdot p_i\epsilon_f^*\cdot p_f\over p_i\cdot k_i}-{\epsilon_i\cdot p_f\epsilon_f^*\cdot p_i\over p_i\cdot k_f}-\epsilon_i\cdot\epsilon_f^*\right)\nonumber\\
&\times&\left({\epsilon_i\cdot p_i\epsilon_f^*\cdot p_f\over p_i\cdot k_i}-{\epsilon_i\cdot p_f\epsilon_f^*\cdot p_i\over p_i\cdot k_f}-\epsilon_i\cdot\epsilon_f^*\right)
\end{eqnarray}
which may be checked (with a bit of effort).

Again it is convenient to use the helicity formalism.  Using factorization, the center-of-mass helicity amplitudes for spin-0 gravitational Compton annihilation of a particle with mass $m_A$ are found to be
\begin{eqnarray}
^AB_0^{grav}(++)&=&^AB_0^{grav}(--)={\kappa^2\over 4}(E^2-\boldsymbol{p}_A^2x^2)[{m_A^2\over E^2-\boldsymbol{p}_A^2x_A^2}]^2={\kappa^2\over 4}{m_A^4\over E^2-\boldsymbol{p}_A^2x_A^2}\,,\nonumber\\
^AB_0^{grav}(+-)&=&^AB_0^{grav}(-+)={\kappa^2\over 4}(E^2-\boldsymbol{p}_A^2x_A^2)[{\boldsymbol{p}_A^2(1-x_A^2)\over E^2-\boldsymbol{p}_A^2x_A^2}]^2={\kappa^2\over 4}{\boldsymbol{p}_A^4(1-x_A^2)^2\over E^2-\boldsymbol{p}_A^2x_A^2} \,.\nonumber\\
\quad
\end{eqnarray}
Making the analytic continuation, as before, we find
\begin{eqnarray}
^AB_0^{grav}(++)&=&^AB_0^{grav}(--)\stackrel{FS}{\longrightarrow}{\kappa^2m_A^2\xi_A^2\over 4}{(1+\tau_A^2)^2\over d_A}\nonumber\\
^AB_0^{grav}(+-)&=&^AB_0^{grav}(-+)\stackrel{FS}{\longrightarrow}{\kappa^2m_A^2\xi_A^2\over 4}{(1-x_A^2)^2\over d_A}
\end{eqnarray}
and similar forms hold for the final state reaction $\gamma_1+\gamma_2\rightarrow P+P'$ wherein two photons annihilate into a pair of spinless particles with mass $m_B$.
We can write then the succinct forms
\begin{eqnarray}
^AB_0^{grav}(ij)&=&{\kappa^2m_A^2\xi_A^2d_A\over 4}{\cal O}_A^{rs}\epsilon^{*i}_r\epsilon^{*j}_s{\cal O}_A^{uv}\epsilon^{*i}_u\epsilon^{*j}_v\nonumber\\
^BB_0^{grav}(ij)&=&{\kappa^2m_B^2\xi_B^2d_B\over 4}\epsilon^{i}_r\epsilon^{j}_s{\cal O}_B^{rs}\epsilon^{i}_u\epsilon^{j}_v{\cal O}_B^{uv}
\end{eqnarray}
and substituting in Eq. (\ref{eq:mk}), determine the discontinuity for the gravitational scattering of spinless particles having masses $m_A,m_B$ across the two-graviton t-channel cut, {\it cf.} Figure 1\footnote{Note that we have again divided by the normalizing factor $4m_Am_B$.}
\begin{eqnarray}
&&{\rm Disc}\,{\rm Amp}_2^{grav}(q)=-{i\over 2!}{1\over 4m_Am_B}\int {d^3k_1\over (2\pi)^32k_1^0}{d^3k_2\over (2\pi)^32k_2^0}(2\pi)^4\delta^4(p_1+p_2-k_1-k_2)\nonumber\\
&\times& \sum_{a,b=1}^2{\kappa^4m_A^2\xi_A^2m_B^2\xi_B^2d_Ad_B\over 16}{\cal O}_A^{ij}\epsilon^{a*}_{1i}\epsilon^{b*}_{2j}{\cal O}_A^{k\ell}\epsilon^{a*}_{1k}\epsilon^{b*}_{2\ell}\epsilon^a_{1a}\epsilon^b_{2b}{\cal O}_B^{ab*}\epsilon^a_{1c}\epsilon^b_{2\ell}{\cal O}_B^{cd*}\nonumber\\
&=&-i{\kappa^4m_A\xi_A^2m_B\xi_B^2\over 1024\pi}<\sum_{i,j,k,\ell=1}^3\sum_{a,b,c,d=1}^3 d_Ad_B{\cal O}_A^{ij}{\cal O}_A^{k\ell}P^G_{ik;ac}P^G_{j\ell;bd}{\cal O}_B^{ab*}{\cal O}_B^{cd*}>\nonumber\\
\quad
\end{eqnarray}
where, as before, $\tau_i=\sqrt{t}/2m_i$, $x_i=\hat{\boldsymbol{p}}_i\cdot\hat{\boldsymbol{k}}$, $d_i=\tau_i^2+x_i^2$, with $i=A,B$ and
$$<G>\equiv \int{d\Omega_{\hat{\boldsymbol{k}}}\over 4\pi}\,G\,,$$
while the sum over graviton polarizations is
\begin{equation}
P^G_{ik;ac}=\sum_{r=1}^2\epsilon^{r*}_i\epsilon^{r*}_i\epsilon^{r}_a\epsilon^{r}_c={1\over 2}\left[\delta^T_{ia}\delta^T_{kc}+\delta^T_{ic}\delta^T_{ka}-\delta^T_{ik}\delta^T_{ac}\right]
\end{equation}
Performing the polarization sum, we find the exact form
\begin{eqnarray}
&&<\left[d_Ad_B{\cal O}_A^{ij}{\cal O}_A^{k\ell}P^G_{ik;ac}P^G_{j\ell;bd}{\cal O}_B^{ab*}{\cal O}_B^{cd*}\right]>=
<{1\over d_Ad_B}\left[4\left(2(y-x_Ax_B)^2\right.\right.\nonumber\\
&-&\left.\left.(1-x_A^2)(1-x_B^2)\right)^2-2(1-x_A^2)^2(1-x_B^2)^2+2(1+\tau_A^2)^2(1+\tau_B^2)^2\right]\nonumber\\
\quad
\end{eqnarray}

Using the angular integrals $I_{mn}$ from Appendix B, we find then
\begin{eqnarray}
&&{\rm Disc}\,{\rm Amp}^{grav}_2(q)\stackrel{t<<m_A^2,m_B^2}{\longrightarrow}-i{\kappa^4m_A\xi_A^2m_B\xi_B^2\over 512\pi}<{1\over d_Ad_B}\left[
8(y-x_Ax_B)^4\right.\nonumber\\
&-&\left.8(y-x_Ax_B)^2(1-x_A^2)(1-x_B)^2+(1-x_A^2)^2(1-x_B^2)^2+1\right]>\nonumber\\
&&\stackrel{y\rightarrow 1}{\longrightarrow}-i{\kappa^4m_A\xi_A^2m_B\xi_B^2\over 512\pi}<{1\over d_Ad_B}\left[ x_A^4x_B^4+6x_A^4x_B^2+6x_A^2x_B^4-16x_A^3x_B^3\right.\nonumber\\
&-&\left.16x_A^3x_B-16x_Ax_B^3+x_A^4+x_B^4+36x_A^2x_B^2+6x_A^2\right.\nonumber\\
&+&\left.6x_B^2-16x_Ax_B+2\right]\nonumber\\
&=&-i{\kappa^4m_A\xi_A^2m_B\xi_B^2\over 512\pi}\left[I_{44}+6I_{42}+6I_{24}-16I_{33}-16I_{13}-16I_{31}\right.\nonumber\\
&+&\left.I_{40}+I_{04}+36I_{22}+6I_{20}+6I_{02}-16I_{11}+2I_{00}\right]\nonumber\\
&=&-i{\kappa^4m_A^2m_B^2\over 128\pi}\left[{1\over 5}+{1\over 3}(6+6-16)+1\cdot(-16-16+1+1+36)\right.\nonumber\\
&+&\left.6\left({\pi m_A\over \sqrt{t}}-1\right)+6\left({\pi m_A\over \sqrt{t}}-1\right)-16\left(-1+i4\pi{m_r\over p_0}\right)\right.\nonumber\\
&+&\left.2\left(-{1\over 3}+i4\pi{m_rm_Am_B\over p_0t}\right)\right]\nonumber\\
&=&-i{\kappa^4m_A^2m_B^2\over 128\pi}\left[{41\over 5}+6{\pi(m_A+m_B)\over \sqrt{t}}+4\pi i{m_Am_Bm_r\over p_0t}+\ldots\right],\label{eq:mg}
\end{eqnarray}
so that the gravitational scattering amplitude is
\begin{equation}
{\rm Amp}_2^{grav}(q)=-{\kappa^4m_Am_B\over 1024\pi^2}\left[{41\over 5}L-6S(m_A+m_B)+4\pi i{m_Am_Bm_r\over p_0t}L+\ldots\right]
\end{equation}

As in the electromagnetic case, the imaginary term is associated with the scattering phase or equivalently the Born iteration, which must be subtracted in order to generate a properly defined second order potential
\begin{eqnarray}
B_2^{grav}(q)&=&i\int{d^3\ell\over (2\pi)^3}{{\kappa^2\over 8}m_A^2\over |\boldsymbol{p}_f-\boldsymbol{\ell}|^2+\lambda^2}
{i\over {p_0^2\over 2m_r}-{\ell^2\over 2m_r}+i\epsilon}{{\kappa^2\over 8}m_B^2\over |\boldsymbol{\ell}-\boldsymbol{p}_i|^2+\lambda^2}\nonumber\\
&=&-i{\kappa^4\over 256\pi}m_A^2m_B^2{m_r\over p_0t}L
\end{eqnarray}
The result is a well-defined second order gravitational potential
\begin{eqnarray}
V_2^{grav}(r)&=&-\int{d^3q\over (2\pi)^3}e^{-i\boldsymbol{q}\cdot\boldsymbol{r}}\left({\rm Amp}_2^{grav}(q)-B_2^{grav}(q)\right)\nonumber\\
&=&-{3Gm_Am_B(m_A+m_B)\over r^2}-{41G^2m_Am_B\hbar\over 10\pi r^3}
\end{eqnarray}
which agrees with the result calculated via Feynman diagram methods by Khriplovich and Kirilin and Bjerrum-Bohr et al.~\cite{Khr03},\cite{Bje03}.  However, use of factorization has allowed an {\it enormously} simpler evaluation. Although identical results for gravitational scattering are obtained from either procedure, as shown in Appendix A, the on-shell technique is calculationally much simpler since the Feynman diagram procedure involves
\begin{itemize}
\item[a)] not only the bubble, triangle, box, and cross-box diagrams considered in the electromagnetic case and shown in Figure 3 (but now with tensor vertices associated with the energy-momentum tensor replacing the vector vertices associated with the electromagnetic current and fourth-rank tensor graviton propagators replacing second rank tensor photon propagators), but also

\item[b)] completely new 5a),5b) vertex-bubble and 5c),5d) vertex-triangle diagrams involving the sixth-rank tensor triple graviton vertex, together with the vacuum polarization diagram, which involves {\it two} triple graviton vertices, as shown in Figure 5e).  In addition, the vacuum polarization contribution must be modified by subtracting the ghost loop diagram shown in Figure 5f).
\end{itemize}
It should also be emphasized that since the discontinuity involves only {\it physical} states, it is unnecessary in the on-shell method to include any contributions involving ghosts.

\begin{figure}
\begin{center}
\epsfig{file=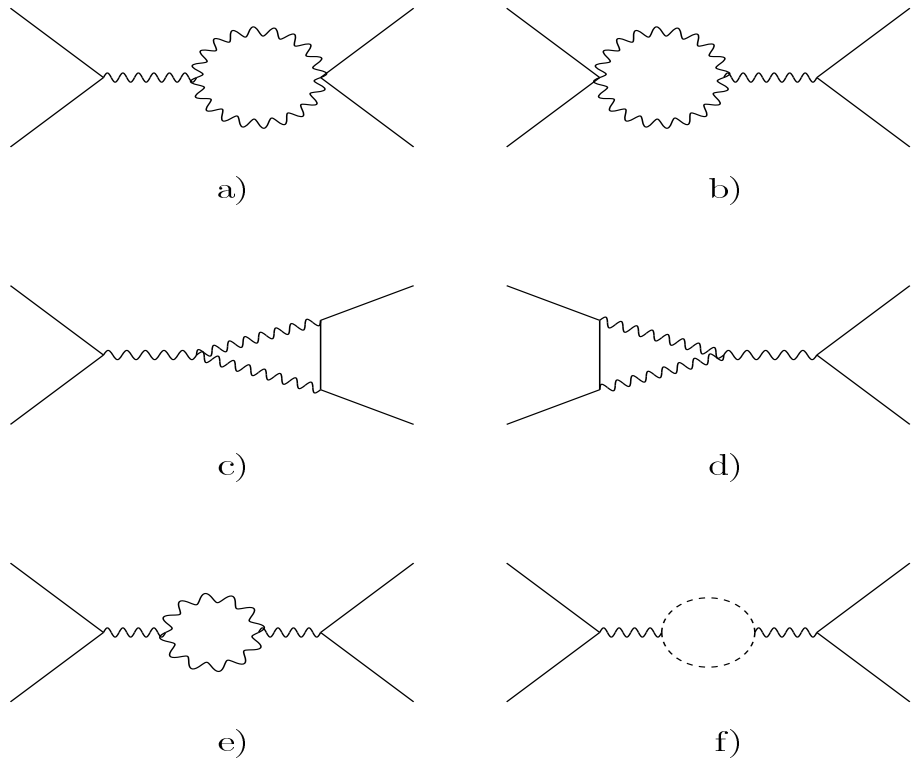,height=10cm,width=12cm}
\caption{{Shown are the a),b) vertex-bubble, c),d) vertex-triangle, e) vacuum polarization and f) ghost-loop diagrams contributing to spinless particle gravitational scattering, all but the ghost-loop involving the triple-graviton vertex. Here the solid lines represent the massive spinless particles, while the wiggly lines represent gravitons.}}
\end{center}
\end{figure}

\subsection{Polarizable Gravitational Scattering}

Before concluding, we consider an extension of gravitational scattering to the situation that the spinless system $A$ is polarizable.  Of course, in the gravitational case this polarization takes the form of a quadrupole, rather than a dipole as found in the electromagnetic case, and leads to a bit more complication.  The quadrupole arises due to the presence of a gravitational field gradient, and we define the gravitational polarizability $\alpha_G$ as the constant of proportionality between the applied field gradient and the induced dipole moment, via~\cite{For15}
\begin{equation}
Q_{ij}\equiv \alpha_G^A R_{0i;0j}\label{eq:gd}
\end{equation}
where we have expressed the field gradient in terms of the Riemann curvature tensor.  There will exist a classical potential if the field is generated by a mass $m_B$ located at the origin, in which case, if the polarizable system is located at position $\boldsymbol{r}$, the local curvature is
\begin{equation}
R_{0i;0j}={1\over 2}\nabla_i\nabla_j h_{00}={}-{1\over 32\pi}\left(3{r_ir_j\over r^5}-\delta_{ij}{1\over r^3}\right)\kappa m_B
\end{equation}
so that the induced quadrupole moment becomes
\begin{equation}
Q_{ij}=\alpha_G^A{\kappa m_B\over 32\pi}\left(3{r_ir_j\over r^5}-\delta_{ij}{1\over r^3}\right)\,,
\end{equation}
leading to the classical potential
\begin{equation}
U(r)=-{1\over 2}\sum_{ij}Q_{ij}R_{0i;0j}=-{1\over 2}\alpha_G^A\sum_{ij}R_{0i;0j}^2=-3\alpha_G^A {Gm_B^2\over 32\pi r^6}
\end{equation}
Alternatively, we can derive this classical form together with its quantum mechanical corrections by finding the amplitude for the creation of a pair of gravitons by a spin-zero system having mass $m_A$.  Using
\begin{equation}
R_{0i0j}={1\over m_A^2}p_1^\alpha p_1^\gamma R_{\alpha i;\gamma j}
\end{equation}
we find the Hamiltonian
\begin{equation}
H=-{1\over 2}\alpha_G^A\sum_{ij}R^2_{0i;0j}=-{\alpha_G^A\over 2m_A^4}p_1^\alpha p_1^\gamma R_{\alpha\beta;\gamma\delta}R^{\rho\beta;\sigma\delta}p_{1\rho}p_{1\sigma}
\end{equation}
Using the linearized approximation for the curvature
\begin{equation}
R_{\alpha\beta;\gamma\delta}=\frac{1}{2}\left( \frac{\partial^2h_{\alpha\delta}}{\partial x^\beta \partial x^\gamma} + \frac{\partial^2h_{\beta\gamma}}{\partial x^\alpha \partial x^\delta} - \frac{\partial^2h_{\alpha\gamma}}{\partial x^\beta \partial x^\delta} - \frac{\partial^2h_{\beta\delta}}{\partial x^\alpha \partial x^\gamma} \right)
\end{equation}
we determine the amplitude for annihilation into a graviton pair to be
\begin{equation}
{\rm Amp_G}=-{1\over 4}\alpha_G^A\epsilon_{1\alpha}^*\epsilon_{1\beta}^*\epsilon_{2\gamma}^*\epsilon_{2\delta}^*T_Q^{\alpha\beta;\gamma\delta}(p_1,k_1,k_2)
\end{equation}
where
\begin{eqnarray}
T_Q^{\alpha\beta;\gamma\delta}(p_1,k_1,k_2)&=&{1\over m_A^4}\left(p_1^\alpha p_1\cdot k_1(k_{1\rho}\eta^\beta_\sigma+\eta^\beta_\rho k_{1\sigma})-p_1^\alpha p_1^\beta k_{1\rho} k_{1\sigma}-(p_1\cdot k_1)^2\eta^\alpha_\rho\eta^\beta_\sigma\right)\nonumber\\
&\times&\left(p_1^\gamma p_1\cdot k_2(k_2^\rho\eta^{\delta\sigma}+\eta^{\delta\rho} k_2^\sigma)-p_1^\gamma p_1^\delta k_2^\rho k_2^\sigma-(p_1\cdot k_2)^2\eta^{\gamma\rho}\eta^{\delta\sigma}\right)
\end{eqnarray}
is the quadrupole transition amplitude.  On account of the symmetries of $R_{\alpha\beta;\gamma\delta}$, the quadrupole amplitude factorizes and can be written in terms of a product of electric transition amplitudes
\begin{equation}
T_Q^{\alpha\beta;\gamma\delta}(p_1,k_1,k_2)=T_E^{\alpha\gamma}(p_1,k_1,k_2)T_E^{\beta\delta}(p_1,k_2,k_2)
\end{equation}
The quadrupole two-graviton annihilation amplitude is then given as a product of dipole electric transition amplitudes
\begin{equation}
{\rm Amp}_G=-{1\over 4}\alpha_G^A\times\left({\rm Amp}_E\over 4\pi\alpha_E\right)^2
\end{equation}
so that the corresponding helicity amplitudes factorize
\begin{eqnarray}
{\rm Amp}_G(++)&=&{\rm Amp}_G(--)=-{1\over 4}\alpha_G^A\times\left({\rm Amp}_E(++)\over 4\pi\alpha_E\right)^2=-\alpha_G^A{t^2\over 64}\nonumber\\
{\rm Amp}_G(+-)&=&{\rm Amp}_G(-+)=-{1\over 4}\alpha_G^A\times\left({\rm Amp}_E(+-)\over 4\pi\alpha_E\right)^2=-\alpha_G^A{t^2\over 64}(1-x_A^2)^2\nonumber\quad
\end{eqnarray}
The contact helicity amplitude for emission of a graviton pair can be written as
\begin{equation}
S^{ab}_A= -t^2{\cal U}^{ij;kl}\epsilon^{a*}_{1i}\epsilon^{a*}_{1k}\epsilon^{b*}_{2j}\epsilon^{b*}_{2\ell}
\end{equation}
with
\begin{equation}
{\cal U}_A^{ij;k\ell}=-{\alpha_G^A\over 64}d_A^2{\cal O}_A^{ij}{\cal O}_A^{k\ell}
\end{equation}
For the case of a point mass $m_B$ interacting with a polarizable mass $m_A$, we have then
\begin{eqnarray}
&&{\rm Disc}\,{\rm Amp}_2^{N-grav}\,=-{i\over 2!}\int {d^3k_1\over (2\pi)^32k_1^0}{d^3k_2\over (2\pi)^32k_2^0}(2\pi)^4\delta^4(p_1+p_2-k_1-k_2)\nonumber\\
&\times&\sum_{a,b=1}^2{t^2\kappa^2m_B^2\xi_B^2d_B\over 8m_B}{\cal U}_A^{ij;k\ell}\epsilon^{a*}_{1i}\epsilon^{a*}_{1k}\epsilon^{b*}_{2j}\epsilon^{b*}_{2\ell}\epsilon^a_{1u}\epsilon^a_{1v}\epsilon^b_{2r}\epsilon^b_{2s}{\cal O}_B^{ur*}{\cal O}_B^{vs*}\nonumber\\
&=&{-i\kappa^2m_B\xi_B^2t^2\over 128}
\sum_{i,j,k,\ell=1}^3\sum_{u,v,r,s=1}^3<\left[{\cal U}_A^{ij;k\ell}P^T_{ik;uv}P^T_{j\ell;rs}{\cal O}_B^{ur*}{\cal O}_B^{vs*}\right]d_B>\nonumber\\
&\stackrel{y\rightarrow 1}{\longrightarrow}&{\kappa^2m_B\alpha_G^At^2\over 4096\pi}\left[J^B_{44}+6J^B_{42}+6J^B_{24}\right.\nonumber\\
&-&\left.16J^B_{33}-16J^B_{13}-16J^B_{31}+J^B_{40}+J^B_{04}+36J^B_{22}+6J^B_{20}+6J^B_{02}-16J^B_{11}+2J^B_{00}\right]\nonumber\\
&=&{\kappa^2m_B\alpha_G^At^2\over 4096\pi}\left[2{\pi m_B\over \sqrt{t}}+1\cdot(-2-16+6+6)+{1\over 3}\cdot(36+1+1-16-16)\right.\nonumber\\
&+&\left.{1\over 5}\cdot(-16+6+6)+{1\over 7}\right]={\kappa^2m_B\alpha_G^At^2\over 4096\pi}\left[{2\pi m_B\over \sqrt{t}}-{163\over 35}\right]
\end{eqnarray}
so
\begin{equation}
{\rm Amp}_2^{N-grav}=-{Gm_B\alpha_G^A\over 256\pi}\left(2m_BSt^2-{163\over 35}t^2L\right)
\end{equation}
Setting $t=q^2$ and taking the Fourier transform, we find the effective potential
\begin{equation}
V_2^{N-grav}(r)=-\int{d^3q\over (2\pi)^3}e^{-i\boldsymbol{q}\cdot\boldsymbol{r}}{\rm Amp}_2^{N-grav}(q)=-{3Gm_B^2\alpha_G^A\over 32\pi}\left({1\over r^6}+{163\over 7}{\hbar\over \pi m_Br^7}\right)\label{eq:fx}
\end{equation}

Finally, in the case of the long-range interaction of a pair of polarizable systems, we have
\begin{eqnarray}
&&{\rm Disc}\,{\rm Amp}_2^{NN-grav}(q)\,=-{i\over 2!}\int {d^3k_1\over (2\pi)^32k_1^0}{d^3k_2\over (2\pi)^32k_2^0}(2\pi)^4\delta^4(p_1+p_2-k_1-k_2)\nonumber\\
&\times&\sum_{a,b=1}^2{t^4}{\cal U}_A^{ij;k\ell}\epsilon^{a*}_{1i}\epsilon^{a*}_{1k}\epsilon^{b*}_{2j}\epsilon^{b*}_{2\ell}\epsilon^a_{1u}\epsilon^a_{1v}\epsilon^b_{2r}\epsilon^b_{2s}{\cal U}_B^{uv;rs}\nonumber\\
&=&{t^4\over 16}\sum_{i,j,k,\ell=1}^3\sum_{u,v,r,s=1}^3<{\cal U}_A^{ik;uv}P^T_{ik;uv}P^T_{j\ell;rs}{\cal U}_B^{uv;rs}>\nonumber\\
&=&-i{\alpha_G^A\alpha_G^Bt^4\over 32768\pi}\left[K_{44}+6K_{42}+6K_{24}-16K_{33}-16K_{13}-16K_{31}+K_{40}+K_{04}\right.\nonumber\\
&+&\left.36K_{22}+6K_{20}+6K_{02}-16K_{11}+2K_{00}\right]\nonumber\\
&=&-i{\alpha_G^A\alpha_G^Bt^4\over 32768\pi}\left[2+{1\over 3}(-16+6+6)+{1\over 5}(36+1+1-16-16)\right.\nonumber\\
&+&\left.{1\over 7}(-16+6+6)+{1\over 9}\right]=-i{\alpha_G^A\alpha_G^Bt^4\over 32768\pi}{443\over 315}
\end{eqnarray}
so
\begin{equation}
{\rm Amp}_2^{NN-grav}(q)\,={\alpha_G^A\alpha_G^Bt^4\over 32768\pi^2}{443\over 630}L
\end{equation}
Taking the Fourier transform, we find the effective potential
\begin{equation}
V_2^{NN-grav}(r)=-\int{d^3q\over (2\pi)^3}e^{-i\boldsymbol{q}\cdot\boldsymbol{r}}{\rm Amp}_2^{NN-grav}(q)=-{3987\over 1024}{\alpha_G^A\alpha_G^B\over \pi^3r^{11}}
\end{equation}
which agrees precisely with the retarded form given in \cite{For15},\cite{Wu16}, when we take into account the difference in the definition of polarizability $\alpha_G$ used in our paper and $\alpha_{1S}$ used in theirs---
\begin{equation}
\alpha_G^A\equiv 16\pi G\alpha_{1S}\quad{\rm and}\quad \alpha_G^B\equiv  16\pi G\alpha_{2S}
\end{equation}

As seen in the case of the electromagnetic interaction, the charged-charged, charged-polarizable, and polarizabole-polarizable calculations are related by changing the overall normalization and the substitution $I_{mn}\rightarrow J^A_{mn}\rightarrow K_{mn}$.

\section{Conclusion}

Above we have shown how use of the dispersive methods first studied by Feinberg and Sucher~\cite{Fei88}, involving evaluation of the discontinuity across the t-channel two-photon cut, combined with helicity techniques provides a direct and relatively simple way to determine the asymptotic (long-range) forms of the higher order potentials which characterize the interaction of both charged and neutral spinless systems.  Results were found in each case to agree with previously determined forms, but were obtained with much less calculational effort. The simplification afforded by the use of on-shell procedures arises essentially due to interchange of the integration and summation steps.  That is, in the traditional Feynman diagram approach, the (four-dimensional) Feynman integration is performed first, yielding a set of (gauge-dependent) diagrams which are summed to produced the final (gauge-independent) form.  However, in the on-shell procedure used above the Compton scattering diagrams are first summed to produce gauge-independent helicity amplitudes, which are then combined to provide, via a (two-dimensional) solid-angle integration, the discontinuity across the right-hand cut. This discontinuity can be inserted into a Feynman integral or can be integrated directly. In either case, the result is a scattering amplitude with the correct nonanalytic structure, and thereby long-range form, but with much less effort.  By generalizing to the two-graviton cut, the procedure was also used to calculate the higher order gravitational interaction, where again results found using these methods were shown to agree completely with forms previously calculated using much more complex Feynman diagram methods. In the gravitational case, the use of factorization allows a representation in terms of the product of simple electromagnetic Compton scattering amplitudes, and in addition, the use of these on-shell techniques permits the neglect of any consideration of ghost terms as well as the calculationally cumbersome graviton pole diagrams.

The evaluation of the long-range behavior of the interaction of massless scalar systems has then been shown to be straightforwardly and completely accomplished by the use of on-shell dispersive methods.  This technique can also be successfully applied to the case where one of the scattering systems is massless, as described in a companion paper elucidating this case and containing calculational details for both massive and massless systems~\cite{Hol16}.  A challenge for the future is to understand what happens if one or both of the scattering systems carries spin and/or if {\it both} become massless.  These subjects are under investigation.

\begin{center}
{\bf\large Appendix A}
\end{center}

The treatment of spinless electromagnetic and gravitational scattering given in the above discussion is based on unitarity and is an example of the use of the simplicities which rise from the use of on-shell methods.  It is similar, but easier to implement than the parallel analysis recently presented by Bjerrum-Bohr et al.~\cite{Bje14}, and in this Appendix we demonstrate the complete equivalence of the two analyses.  In the language of \cite{Bje14} the scattering amplitude is constructed in terms of singlet (same helicity) and non-singlet (opposite helicity) intermediate state contributions to the discontinuity, which are inserted into a covariant Feynman integral in order to yield the desired loop amplitude, after which the low energy limit is taken.  In the case of the electromagnetic interaction and spinless scattering
\begin{equation}
{\rm Amp}_2^{em}(q)={i\over 2!}{e^4\over 2}\int {d^4\ell\over (2\pi)^4}{d^4\ell'\over (2\pi)^4}(2\pi)^4\delta^4(p_1+p_2-\ell-\ell'){{\cal N}_{em}\over \ell^2\ell^{'2}\prod_{i=1}^4 p_i\cdot\ell}
\end{equation}
with
\begin{equation}
{\cal N}^{++}_{em}=m_A^2\xi_A^2m_B^2\xi_B^2t^2\quad{\rm and}\quad {\cal N}_{em}^{+-}={\cal E}^2-4{\cal O}
\end{equation}
and
\begin{eqnarray}
{\cal E}&=&2(\ell\cdot p_1\ell\cdot p_4+\ell\cdot p_3\ell\cdot p_2-\ell\cdot\ell'p_1\cdot p_3)\nonumber\\
{\cal O}&=&(\epsilon^{\alpha\beta\gamma\delta}\ell_\alpha p_{1\beta}\ell'_{\gamma}p_{3\delta})^2
\end{eqnarray}
In \cite{Bje14} the scattering amplitude is then evaluated by performing the covariant Feynman integrals directly diagram by diagram, simplifying via the use of Ward identities when possible.

In order to compare with our result we note that, according to the Cutkosky rules, the discontinuity of the scattering amplitude across the two photon cut is
\begin{eqnarray}
{\rm Disc}\,{\rm Amp}_2^{em}(q)&=&-{i\over 2!}{e^4\over 2}\int {d^3\ell\over (2\pi)^32\ell_0}{d^3\ell'\over (2\pi)^32\ell'_0}(2\pi)^4\delta^4(p_1+p_2-\ell-\ell'){{\cal N}_{em}\over \prod_{i=1}^4p_i\cdot\ell}\nonumber\\
&=&-i{e^4\over 32\pi}<{{\cal N}_{em}\over \prod_{i=1}^4p_i\cdot\ell}>
\end{eqnarray}
Performing the Feinberg-Sucher continuation, we have
\begin{eqnarray}
{\cal E}&=&\stackrel{t<<m_A^2,m_B^2}{\longrightarrow}m_A\xi_Am_B\xi_Bt(y-x_Ax_B)\nonumber\\
4{\cal O}&=&\stackrel{t<<m_A^2m_B^2}{\longrightarrow}m_A^2\xi_A^2m_B^2\xi_B^2t^2(1-x_A^2)(1-x_B^2)\sin^2\phi\nonumber\\
&=&m_A^2\xi_A^2m_B^2\xi_B^2t^2(1-x_A^2)(1-x_B^2)(1-\cos^2\phi)\nonumber\\
&=&m_A^2\xi_A^2m_B^2\xi_B^2t^2\left[(1-x_A^2)(1-x_B^2)-(y-x_Ax_B)^2\right]
\end{eqnarray}
where we have chosen
\begin{equation}
\hat{\ell}=\hat{z}\quad \hat{p}_1=\hat{z}x_A+\hat{x}\sqrt{1-x_A^2}\quad \hat{p}_3=\hat{z}x_B+\hat{x}\sqrt{1-x_B^2}\cos\phi+\hat{y}\sqrt{1-x_B^2}\sin\phi
\end{equation}
so
\begin{equation}
y=\hat{p}_A\cdot\hat{p}_B=x_Ax_B+\sqrt{1-x_A^2}\sqrt{1-x_B^2}\cos\phi
\end{equation}
Then
\begin{equation}
{\cal E}^2-4{\cal O}=m_A^2\xi_A^2m_B^2\xi_B^2t^2\left[2(y-x_Ax_B)^2-(1-x_A^2)(1-x_B^2)\right]
\end{equation}
and
\begin{eqnarray}
{\cal N}^{tot}_{em}&=&m_A^2\xi_A^2m_B^2\xi_B^2t^2\left[1+{\cal E}^2-4{\cal O}\right]=m_A^2m_B^2t^2\left[1+2(y-x_Ax_B)^2-(1-x_A^2)(1-x_B^2)\right]\nonumber\\
&=&16E^4m_A^2\xi_A^2m_B^2\xi_B^2\left[2y^2-4yx_Ax_B+x_A^2+x_B^2+x_A^2x_B^2\right]
\end{eqnarray}
Using
\begin{equation}
D\equiv\prod_{i=1}^4p_i\cdot\ell=E^4(E^2-\boldsymbol{p}_A^2x_A^2)(E^2-\boldsymbol{p}_B^2x_B^2)\stackrel{FS}{\longrightarrow}E^4m_A^2\xi_A^2m_B^2\xi_B^2d_Ad_B
\end{equation}
we have
\begin{equation}
{{\cal N}^{tot}_{em}\over D}=16{2y^2-4yx_Ax_B+x_A^2+x_B^2+x_A^2x_B^2\over d_Ad_B}
\end{equation}
and
\begin{equation}
{\rm Disc}\,{\rm Amp}_2^{em}(q)=-i{e^4\over 32\pi}<{{\cal N}^{tot}_{em}\over D}>=-i{e^4\over 2\pi}<{2y^2-4yx_Ax_B+x_A^2+x_B^2+x_A^2x_B^2\over d_Ad_B}>
\end{equation}
in agreement with Eq. (\ref{eq:gb}).

Likewise in the gravitational case
\begin{equation}
{\rm Amp}_2^{grav}(q)=-{i\over 2!}{\kappa^4\over 16}\int {d^4\ell\over (2\pi)^4}{{\cal N}_{grav}\over \ell^2(\ell-q)^2\prod_{i=1}^4p_i\cdot\ell}
\end{equation}
and
\begin{eqnarray}
{\rm Disc}\,{\rm Amp}^{grav}_2(q)&=&-{i\over 2!}{\kappa^4\over 16}\int {d^3\ell\over (2\pi)^32\ell_0}{d^3\ell'\over (2\pi)^32\ell'_0}(2\pi)^4\delta^4(p_1+p_2-\ell-\ell'){{\cal N}_{grav}\over \prod_{i=1}^4p_i\cdot\ell}\nonumber\\
&=&-i{\kappa^4\over 256\pi}<{{\cal N}_{grav}\over \prod_{i=1}^4p_i\cdot\ell}>
\end{eqnarray}
In this case
\begin{equation}
{\cal N}^{++}_{grav}={1\over 8}t^2m_A^4\xi_A^4m_B^4\xi_B^4
\end{equation}
and
\begin{equation}
{\cal N}^{+-}_{grav}={1\over 8t^2}\left[({\cal E}^2-4{\cal O})^2-16{\cal E}^2{\cal O}\right]
\end{equation}
Again, we can check against our results by taking Feinberg-Sucher limit and evaluating the discontinuity.  Since
\begin{equation}
{\cal N}^{++}_{grav}\stackrel{t<<m_A^2,m_B^2}{\longrightarrow} 2E^4m_A^4\xi_A^4m_B^4\xi_B^4
\end{equation}
while
\begin{eqnarray}
&&{\cal N}^{+-}_{grav}\stackrel{t<<m_A^2,m_B^2}{\longrightarrow}2E^4m_A^4\xi_A^4m_B^4\xi_B^4\left[8(y-x_Ax_B)^4-8(y-x_Ax_B)^2(1-x_A^2)(1-x_B^2)\right.\nonumber\\
&+&\left.(1-x_A^2)^2(1-x_B^2)^2\right]
\end{eqnarray}
we have
\begin{eqnarray}
&&{{\cal N}^{tot}_{grav}\over D}=2m_A^2\xi_A^2m_B^2\xi_B^2<{1\over d_Ad_B}\left[8(y-x_Ax_B)^4-8(y-x_Ax_B)^2(1-x_A^2)(1-x_B^2)\right.\nonumber\\
&+&\left.(1-x_A^2)^2(1-x_B^2)^2+1\right]>\nonumber\\
\quad
\end{eqnarray}
Then
\begin{eqnarray}
&&{\rm Disc}\,{\rm Amp}_2^{grav}(q)=-i{\kappa^4\over 128\pi}m_A^2\xi_A^2m_B^2\xi_B^2<{1\over d_Ad_B}\left[8(y-x_Ax_B)^4\right.\nonumber\\
&-&\left.8(y-x_Ax_B)^2(1-x_A^2)(1-x_B^2)+(1-x_A^2)^2(1-x_B^2)^2+1\right]>\nonumber\\
\end{eqnarray}
in agreement with Eq. (\ref{eq:mg}) so that again the scattering amplitude calculated via the two techniques are identical, up to analytic (short-distance) terms.

The essential difference between the two procedures is that \cite{Bje14} does {\it not} directly integrate the discontinuity, but instead inserts it into the Feynman loop integral and then performs the various covariant integrations diagram by diagram using computer evaluation and appropriate simplifying Ward identities, after which approximations appropriate to the low energy limit are performed. Since, as shown here, the discontinuities calculated via the two techniques are the same, the scattering amplitudes are also identical, up to possible analytic (short-distance) corrections, but use of the Feinberg-Sucher results for the angular averages to provide the discontinuity allows a rather more direct and considerably simpler route to the desired scattering amplitudes, especially in the gravitational case.

\begin{center}
{\large\bf Appendix B}
\end{center}

In this Appendix we present the needed solid-angle averaged integrals, which are of three kinds.  The simplest is simply the unweighted solid-angle integrals $K_{mn}$, for which we have
\begin{eqnarray}
K_{00}&=&<1>=1\nonumber\\
K_{11}&=&<x_Ax_B>={1\over 3}y={1\over 3}+\ldots\nonumber\\
K_{20}&=&<x_A^2>={1\over 3}\nonumber\\
K_{02}&=&<x_B^2>={1\over 3}\nonumber\\
K_{22}&=&<x_A^2x_B^2>={1\over 3}-{2\over 15}y^2={1\over 5}+\ldots\nonumber\\
K_{31}&=&<x_A^3x_B>={1\over 5}y={1\over 5}+\ldots\nonumber\\
K_{13}&=&<x_Ax_B^3>={1\over 5}y={1\over 5}+\ldots\nonumber\\
K_{40}&=&<x_A^4>={1\over 5}\nonumber\\
K_{04}&=&<x_B^4>={1\over 5}\nonumber\\
K_{33}&=&<x_A^3x_B^3>={2\over 35}y^3+{3\over 35}y={1\over 7}+\ldots\nonumber\\
K_{42}&=&<x_A^4x_B^2>={4\over 35}y^2+{1\over 35}={1\over 7}+\ldots\nonumber\\
K_{24}&=&<x_A^4x_B^2>={4\over 35}y^2+{1\over 35}={1\over 7}+\ldots\nonumber\\
K_{44}&=&<x_A^4x_B^4>={8\over 315}y^4+{8\over 105}y^2+{1\over 105}={1\over 9}+\ldots
\end{eqnarray}

For the case of a single factor of $d_A$ or $d_B$ in the denominator, the "seed" integral $J^i_{00}$ with $i=A,B$ can be performed directly
\begin{equation}
J^i(00)=<{1\over d_i}>={1\over 4\pi}\int_0^{2\pi}d\phi\int_{-1}^1 {dx_i\over \tau_i^2+x_i^2}={1\over \tau_i}{\rm tan}^{-1}{1\over \tau_i}={1\over \tau_i}\left({\pi\over 2}-{\rm tan}^{-1}\tau_i\right)
\end{equation}
and the generalizations can be found via repeated algebraic iterations using the identity $x_i^2=d_i-\tau_i^2$:
\begin{eqnarray}
J_{00}^A&=&<{1\over d_A}>={1\over \tau_A}{\rm tan}^{-1}{1\over \tau_A}={1\over \tau_A}({\pi\over 2}-{\rm tan}^{-1}\tau_A))=-1+{\pi m_A\over \sqrt{t}}+\ldots\nonumber\\
J_{00}^B&=&<{1\over d_B}>={1\over \tau_B}{\rm tan}^{-1}{1\over \tau_B}={1\over \tau_B}({\pi\over 2}-{\rm tan}^{-1}\tau_B))=-1+{\pi m_B\over \sqrt{t}}+\ldots\nonumber\\
J_{11}^A&=&<{x_Ax_B\over d_A}>=y\left(1-\tau_A^2J_{00}^A\right)=1+\ldots\nonumber\\
J_{11}^B&=&<{x_Ax_B\over d_B}>=y\left(1-\tau_B^2J_{00}^B\right)=1+\ldots\nonumber\\
J_{20}^A&=&<{x_A^2\over d_A}>=1-\tau_A^2J_{00}^A=1+\ldots\nonumber\\
J_{20}^B&=&<{x_B^2\over d_B}>=1-\tau_B^2J_{00}^B=1+\ldots\nonumber\\
J_{02}^A&=&<{x_B^2\over d_A}>={1\over 2}(1-y^2)+{1\over 2}(3y^2-1)J^A_{20}=1+\ldots\nonumber\\
J_{02}^B&=&<{x_A^2\over d_B}>={1\over 2}(1-y^2)+{1\over 2}(3y^2-1)J^B_{20}=1+\ldots\nonumber\\
J_{31}^A&=&<{x_A^3x_B\over d_A}>=y-y\tau_A^2\left(1-\tau_A^2J_{00}^A\right)={1\over 3}+\ldots\nonumber\\
J_{13}^A&=&<{x_Ax_B^3\over d_A}>={y\over 2}[{4\over 3}-\tau_A^2-\tau_A^2(1-\tau_A^2)J_{00}^A]-{y^3\over 2}[{2\over 3}+\tau_A^2-\tau_A^2(1+\tau_A^2)J_{00}^A]={1\over 3}+\ldots\nonumber\\
J_{31}^B&=&<{x_B^3x_A\over d_B}>=y-y\tau_B^2\left(1-\tau_B^2J_{00}^B\right)={1\over 3}+\ldots\nonumber\\
J_{13}^B&=&<{x_Bx_A^3\over d_B}>={y\over 2}[{4\over 3}-\tau_B^2-\tau_B^2(1-\tau_B^2)J_{00}^B]-{y^3\over 2}[{2\over 3}+\tau_B^2-\tau_B^2(1+\tau_B^2)J_{00}^B]={1\over 3}+\ldots\nonumber\\
J_{22}^A&=&<{x_A^2x_B^2\over d_A}>={1\over 3}-\tau_A^2J^A_{02}={1\over 3}+\ldots\nonumber\\
J_{22}^B&=&<{x_A^2x_B^2\over d_B}>={1\over 3}-\tau_B^2J^B_{02}={1\over 3}+\ldots
\end{eqnarray}
where the ellipses indicate terms higher order in $t$.

In the case of integrals involving the product $d_Ad_B$ in the denominator, the seed integrals $I_{00},\,I_{11}$ have been given by Feinberg and Sucher~\cite{Fei70}
\begin{eqnarray}
I_{00}&=&<{1\over d_Ad_B}>={1\over 2\tau_A\tau_B}(F_++{\pi\over N_+})=-{1\over 3}+i2\pi{m_Am_Bm_r\over tp}+\ldots\nonumber\\
I_{11}&=&<{x_Ax_B\over d_Ad_B}>={1\over 2}(F_-+{\pi\over N_+})=-1+i\pi{m_r\over 2p}+\ldots
\end{eqnarray}
where
\begin{equation}
F_\pm(s,t)=\pm{1\over N_-}{\rm tan}^{-1}{N_-\over D_+}-{1\over N_+}{\rm tan}^{-1}{N_+\over D_-}
\end{equation}
with
\begin{eqnarray}
N_\pm(s,t)&=&(\tau_A^2+\tau_B^2+1-y^2\pm 2\tau_A\tau_By)^{1\over 2}\nonumber\\
D_\pm(s,t)&=&y\pm\tau_A\tau_B
\end{eqnarray}
and the generalizations are again found by repeated algebraic iteration
\begin{eqnarray}
I_{20}&=&<{x_A^2\over d_Ad_B}>=J^B_{00}-\tau_A^2I_{00}=-1+{\pi m_B\over \sqrt{t}}-i\pi{m_B\over 2m_A}{m_r\over p}+\ldots\nonumber\\
I_{02}&=&<{x_B^2\over d_Ad_B}>=J^A_{00}-\tau_B^2I_{00}=-1+{\pi m_A\over \sqrt{t}}-i\pi{m_A\over 2m_B}{m_r\over p}+\ldots\nonumber\\
I_{22}&=&<{x_A^2x_B^2\over d_Ad_B}>=1-\tau_A^2J^B_{20}-\tau_B^2J^A_{20}+\tau_A^2\tau_B^2I_{00}=1+\ldots\nonumber\\
I_{31}&=&<{x_A^3x_B\over d_Ad_B}>=J^B_{11}-\tau_A^2I_{11}=1+\ldots\nonumber\\
I_{13}&=&<{x_Ax_B^3\over d_Ad_B}>=J^A_{11}-\tau_B^2I_{11}=1+\ldots\nonumber\\
I_{40}&=&<{x_A^4\over d_Ad_B}>=J^B_{02}-\tau_A^2I_{20}=1+\ldots\nonumber\\
I_{04}&=&<{x_B^4\over d_Ad_B}>=J^A_{02}-\tau_B^2I_{02}=1+\ldots\nonumber\\
I_{42}&=&<{x_A^4x_B^2\over d_Ad_B}>=J^B_{22}-\tau_A^2I_{22}={1\over 3}+\ldots\nonumber\\
I_{24}&=&<{x_A^2x_B^4\over d_Ad_B}>=J^A_{22}-\tau_B^2I_{22}={1\over 3}+\ldots\nonumber\\
I_{33}&=&<{x_A^3x_B^3\over d_Ad_B}>=<x_Ax_B>-\tau_A^2J^A_{11}-\tau_B^2J^B_{11}+\tau_A^2\tau_B^2I_{11}={1\over 3}+\ldots\nonumber\\
I_{44}&=&<{x_A^4x_B^4\over d_Ad_B}>=<x_A^2x_B^2>-\tau_A^2J^A_{22}-\tau_B^2J^B_{22}+\tau_A^2\tau_B^2I_{22}={1\over 5}+\ldots
\end{eqnarray}

\begin{center}
{\bf\large Acknowledgement}
\end{center}

It is a pleasure to acknowledge numerous helpful conversations with John Donoghue, which served to greatly clarify the material discussed above.  This work is supported in part by the National Science Foundation under award NSF PHY11-25915.

\end{document}